\documentclass[a4paper,fleqn]{cas-dc}

\usepackage[authoryear,longnamesfirst]{natbib}

\def\tsc#1{\csdef{#1}{\textsc{\lowercase{#1}}\xspace}}
\tsc{WGM}
\tsc{QE}

\usepackage{placeins}

\usepackage{color}
\usepackage{soul}
\usepackage{capt-of}
\usepackage{float}

\usepackage{csquotes}

\usepackage{graphicx}
\usepackage{dcolumn}
\usepackage{bm}

\usepackage[colorlinks]{hyperref}

\usepackage[compat=1.1.0]{tikz-feynman}
\usepackage{geometry}
\usepackage{csquotes}
\usepackage{tikz}

\begin{document}
\let\WriteBookmarks\relax
\def\floatpagepagefraction{1}
\def\textpagefraction{.001}

\shorttitle{}    

\shortauthors{}  



%

\shorttitle{DESI-Like Hubble Expansion From Staged Symmetry Breaking}

\shortauthors{Hoelscher et al.}

\title[mode=title]
{DESI-Like Hubble Expansion From Staged Symmetry Breaking:
Constraints on Renormalizable Models}

\author[1]{Zachary J. Hoelscher}[orcid=0009-0007-3852-6293]
\cormark[1]
\ead{zachary.j.hoelscher@vanderbilt.edu}

\author[1]{Thomas W. Kephart}[orcid=0000-0001-6414-9590]
\ead{thomas.w.kephart@vanderbilt.edu}

\author[1]{Robert J. Scherrer}[orcid=0000-0002-4464-214X]
\ead{robert.scherrer@vanderbilt.edu}

\author[1,2,3]{Kelly Holley-Bockelmann}[orcid=0000-0003-2227-1322]\ead{k.holley@vanderbilt.edu}

\affiliation[1]{
    organization={Department of Physics and Astronomy,
                  Vanderbilt University},
    city={Nashville},
    postcode={37235},
    state={TN},
    country={USA}
}

\affiliation[2]{
    organization={Department of Life and Physical Sciences,
                  Fisk University},
    city={Nashville},
    postcode={37208},
    state={TN},
    country={USA}
}

\affiliation[3]{
    organization={Space Research Organisation Netherlands (SRON)},
    addressline={Niels Bohrweg 4},
    city={Leiden},
    postcode={2333 CA},
    country={The Netherlands}
}

\cortext[1]{Corresponding author: Zachary J. Hoelscher}

\begin{abstract}
The Dark Energy Spectroscopic Instrument (DESI) second data release shows a moderate preference for dark energy with a time-varying equation of state parameter, suggesting that the standard $\Lambda$CDM model may need to be revised. In particular, DESI favors dark energy whose equation of state parameter can drop below $-1$, violating the null energy condition. Chen and Loeb have recently suggested that this violation may be avoided if a subcomponent of the dark matter possesses a time-dependent equation of state. In this work, we explore the space of possible models, in an effort to find a natural, renormalizable realization. We find that once one requires the model to be renormalizable, a construction that generates the desired expansion history is fairly broadly obstructed. While we are unable to produce a fully natural, renormalizable realization of their ideas, this work is valuable for constraining directions one might pursue for viable theoretical explanations of the DESI results. Our framework is also interesting for incorporating features that may be of interest for model building elsewhere, as our decay cascades enable decays that switch on at particular redshifts while also naturally incorporating a self-interacting dark matter candidate with a velocity-dependent cross section as a consequence of gauge invariance. The second feature is relevant for addressing tensions between $\Lambda$CDM and observations of small-scale structure, particularly the diversity of galactic rotation curves. 
\nocite{*}
\end{abstract}




\maketitle
\section{\label{sec:intro} Introduction}

Since the accelerating expansion of the Universe was first discovered in 1998~\citep{Riess_1998, Perlmutter_1999}, the mysteries of dark energy have intrigued the scientific community. While the simplest explanation for this acceleration is a cosmological constant, others have suggested that dark energy could be explained by the addition of a new scalar field~\citep{Peebles_1987, Ratra_1988, Zlatev_1999}, which could allow for dynamical dark energy~\citep{Caldwell_1998}. The recent Dark Energy Spectroscopic Instrument (DESI) results show up to a 4.2$\sigma$ preference for a dynamical equation of state parameter for dark energy, perhaps motivating alternatives beyond the standard $\Lambda$CDM paradigm~\citep{DESI_I, DESI_II, Lodha_2025}, such as modifications to cold dark matter~\citep{Li_2025, Khoury_2025, Abedin_2025, Kumar_2025, Wang_2025, Yang_2025, Yao_2025, Colgain_2025}. Strikingly, some have even found a 5$\sigma$ preference~\citep{Nunes_2025}. 

The DESI results also prefer a model of dark energy that exhibits phantom behavior, meaning an equation of state parameter that drops below -1~\citep{Lodha_2025}. Such behavior seemingly violates the null energy condition, $T_{\mu \nu} k^\mu k^\nu \geq 0$~\citep{Dabrowski_2008}. 

In an effort to avoid this, recent work proposes that the DESI results could instead be explained by a combination of a cosmological constant and a subcomponent of dark matter with a time-varying equation of state parameter~\citep{Chen_2025}. In this work, they assume that the subcomponent has a density parameter that is around one-fifth of the density parameter for the other matter content. They also assume that its equation of state parameter exhibits step-like changes between constant values over the range -0.5 to 0.5, and they examine a few scenarios for $w(z)$. 

In particular,~\citep{Chen_2025} adopt the following potential $V(\phi)$ to motivate this, where $\phi_1 >> \phi_2 > 0$: 

\begin{equation}
    V(\phi) \propto f(\phi)^6 / (\phi_1^4 + f(\phi)^4),
\end{equation}

\noindent and
\begin{equation}
    f(\phi) = \bigg( \phi_2^{16/3} + |\phi|^{16/3} \bigg)^{1/6} |\phi|^{1/9}.
\end{equation} 
Unfortunately, this potential is non-renormalizable and lacks a clear, natural UV completion (a form of the theory that is valid at high energy scales), which is needed for a fully consistent theory. They emphasize, however, that this is but one possible example. A following paper adopts a different model, though this is again non-renormalizable, so one again has divergences that cannot be removed using a finite number of counterterms~\citep{Braglia_2025}.

In this work, we explore the space of natural, perturbatively renormalizable constructions that have the potential to realize Chen and Loeb’s ideas. These are not hard requirements for a consistent theory, as some non-renormalizable theories are sensible effective field theories, though these restrictions enable us to consider a class of possible models which are interesting if one desires a model that could plausibly have a UV completion. We discuss the problems with various directions, and argue that these concerns could perhaps motivate alternative directions. Our main goal is not to claim to have identified the ideal model, but rather to help constrain the space of viable theoretical explanations for the DESI results.

The framework we consider is additionally interesting for naturally accommodating a potential self-interacting dark matter (SIDM) candidate as a by-product of gauge invariance. SIDM has been proposed to resolve tensions between $\Lambda$CDM and observations of small-scale structure, such as the diversity of rotation curves in dwarf galaxies~\citep{Spergel_2000, Kamada_2017, Ren_2019}. $\Lambda$CDM without baryonic physics predicts cuspy NFW density profiles~\citep{Navarro_1996, Navarro_1997, Ludlow_2013}, yet some dwarf galaxies are observed to have cored profiles~\citep{Spergel_2000, Gentile_2004, Oman_2015, Santos_2017, Santos_2020}. Cores can be produced via baryonic feedback~\citep{Cores_Dwarf_Galaxies, Read_Gilmore_2005, Pontzen_2012,
Governato_et_al_2012, Teyssier_2013, Cintio_2013, Onorbe_2015, Chan_2015, Tollet_2016, Read_2016, Dutton_et_al_2019, 
Cores_Dwarf_Galaxies2, Lazar_2020, Jahn_2023, Azartash_2024}; however, some observed dwarf galaxies have too few stars to easily explain their density profiles through such means~\citep{Almeida_2024}. SIDM alleviates these tensions by flattening the central density profile of the dark matter halo, which is achieved through non-gravitational interactions between particles~\citep{Spergel_2000}. 

Our SIDM candidate consists of a heavy vector field interacting through a scalar mediator. Vectors interacting through a scalar have been considered before as SIDM~\citep{Duch_2018}, though our case is interesting because the SIDM can naturally arise from a minor extension of this framework (which already requires multiple U(1) symmetries) as a by-product of constructing a consistent theory. We find that the interaction cross section is velocity-dependent, which is desirable to allow SIDM to produce cored halos at the galaxy scale while also satisfying constraints from galaxy clusters~\citep{Sagunski_2021, Correa_2021, Correa_2022}. 

\section{Discussion of Possible Models}

We wish to produce a dark matter model whose equation of state parameter shifts between four values: $w_1, w_2, w_3,$ and $w_4$. To do this, we consider a decay cascade of the form $A \to B \to C \to D$. We consider a decay chain, rather than a convoluted potential for a single scalar field (similar to that used by~\citep{Chen_2025}), because the desire for a renormalizable model constrains us to only consider interaction terms with a mass dimension of four or less. The crucial point is that this requirement of renormalizability strongly constrains the potentials one can construct, leading one to use some form of decay cascade in this context, if one wishes to obtain the desired shifts in the equation of state parameter. The challenges faced when attempting to construct such a model are then fairly generic, which we discuss in detail in Section~\ref{sec:Broad_Obstruction}. This cascade concept is similar to the dynamical dark matter considered by Dienes~\citep{Dienes_2012, Dienes_2012_b, Dienes_2013}, though our work contains a finite number of stages, rather than an infinite number, and is motivated by DESI observations.

\subsection{Four Real Scalar Fields}
The simplest construction one might try would be a decay cascade involving four real scalar fields ($A$, $B$, $C$, and $D$), similar to that shown below: 

\begin{align}
\mathcal{L} \supset 
& \sum_{\Phi \in A, B, C, D}
  \bigg[
    \frac{1}{2} \partial^\mu \Phi \, \partial_\mu \Phi
    - \frac{1}{2} m_{\Phi}^2 \Phi^2
    - \lambda_{\Phi} \Phi^4
  \bigg] \notag \\[4pt]
& - g_{AB}\, AB^2
  - g_{BC}\, BC^2
  - g_{CD} CD^2 ,
\end{align}

\noindent where this model permits a decay cascade $A \to B \to C \to D$ via tree-level processes such as $A \to 2B$. (Note that a \enquote{tree-level} process is the lowest-order perturbative approximation for a decay or scattering process.) 

Unfortunately, unwanted decay processes will occur at one-loop via diagrams like the one shown below, which yields $A \to 2C$. Such diagrams would allow the decay chain to skip stages, and are still undesirable, although they are suppressed by extra factors of coupling  constants.   

\begin{center}
\begin{tikzpicture}
  \useasboundingbox (-1.2,-0.6) rectangle (2.2,0.6);
  
  \coordinate (A) at (-1,0);
  \coordinate (L) at (0,0);
  \coordinate (T) at (1.0,0.5);
  \coordinate (B) at (1.0,-0.5);
  \coordinate (C1) at (2,0.5);
  \coordinate (C2) at (2,-0.5);

  \fill (L) circle (2pt);
  \fill (T) circle (2pt);
  \fill (B) circle (2pt);

  \draw (A) -- (L);
  \draw (L) -- (T) -- (B) -- (L);
  \draw (T) -- (B);
  \draw (T) -- (C1);
  \draw (B) -- (C2);

  \node[left] at (A) {$A$};
  \node[right] at (C1) {$C$};
  \node[right] at (C2) {$C$};
  \node[above left] at (T) {$B$};
  \node[below left] at (B) {$B$};

  \node[right] at ($(T)!0.5!(B)$) {$C$};
\end{tikzpicture}
\end{center}

Far more importantly, such a model also results in a suppressed effect on the Hubble parameter $H(z)$ because it does not enforce a clean shift of energy density from one field to another. In effect,  while $A$ is decaying to $B$, $B$ is decaying to $C$, and $C$ to $D$, so at any given time, there is a mixture of the four fields, rather than a clean sequential shift of the energy density from one field to another.

\newpage 

\subsection{Four Complex Scalar Fields With Global Symmetry}

We can ensure a clean shift of energy density from one field to another (so the effect on $H(z)$ is not suppressed) by using complex scalars with appropriate charges. One could spontaneously break a global $\rm{U}(1) \times \rm{U}(1) \times \rm{U}(1)$ symmetry to sequentially turn on two-body decays of the form $A \to B^\dagger B$. Sequentially turning on decays in this way helps to ensure a clean shift of energy density between fields. 

Such a model still has an issue, however. Global symmetries are believed to be inconsistent once embedded in a full theory of quantum gravity~\citep{Harlow_2019}. Even without appealing to quantum gravity consistency, one can argue that gauge symmetries are natural. Global symmetries are expected to be only approximate at best, with gauge symmetries taking primacy~\citep{Witten_2018}. This conjecture can be motivated by the AdS/CFT correspondence~\citep{Harlow_2019}, which can be regarded as a toy model for quantum gravity where anti-de Sitter space (AdS) is dual to a conformal field theory (CFT)~\citep{Maldacena_1999}. Our Universe is not anti-de Sitter, as the cosmological constant is believed to be positive, rather than negative, though the arguments used for AdS/CFT could perhaps be generalized to flat-space holography, which would describe a Universe with a vanishing cosmological constant, and which would be a more useful approximate model for our Universe. 

When considering continuous global symmetries, the case that they are inconsistent can also be motivated by the black hole information paradox, in which case the argument can be generalized beyond the AdS/CFT correspondence~\citep{Harlow_2021}. One can loosely motivate this by arguing that continuous global symmetries imply conserved charges that are destroyed by black hole evaporation; therefore, continuous global symmetries should not exist in nature, as we expect black hole evaporation to be unitary. This argument is more restrictive than the AdS/CFT result, as it applies to continuous global symmetries, such as U(1), but not global symmetries in general. One can further motivate this using the Bekenstein-Hawking entropy by appealing to the argument given in the introduction of~\cite{Geng_2026}, though one should note that the global symmetries they later construct in their work are not relevant to astrophysical black holes, which do not incorporate a non-gravitational bath.

\subsection{Four Complex Scalar Fields With Local Symmetry}

We have four complex scalar fields $A$, $B$, $C$, and $D$, with four dark Higgs fields $H_1$, $H_2$, $H_3$, and $H_4$. One might initially suspect that a scalar field will not clump as dark matter because the kinetic terms could imply a sound speed of $c$, though there is no such problem~\citep{Chen_2025}. 
Since the theory has a U(1) $\times$ U(1) $\times$ U(1) $\times$ U(1) gauge group, we must introduce four gauge bosons to allow for gauge invariance: $\alpha_\mu$, $\beta_\mu$, $\gamma_\mu$, and $\delta_\mu$, where we define the Faraday tensors as $\Psi_{\mu \nu} = \partial_\mu \Psi_\nu - \partial_\nu \Psi_\mu$, and the covariant derivative as $\mathcal{D}_\mu \equiv \partial_\mu - i(g_1 q_1 \alpha_\mu + g_2 q_2 \beta_\mu + g_3 q_3 \gamma_\mu + g_4 q_4 \delta_\mu)$. We note that a similar gauge group, U(1) $\times$ U(1) $\times$ U(1), can arise in string theory~\citep{Cvetic_2022}. We also note that anomalies will not be present, as the model does not incorporate fermions~\citep{Feruglio_2021}. In the Lagrangian below, we include generic potentials, e.g. $V(A, A^{\dagger})$, to keep the model general.  

\begin{equation}
\begin{aligned}
\mathcal{L} \supset &
\; \sum_{\Psi_\mu \in \{ \alpha_\mu, \beta_\mu, \gamma_\mu, \delta_\mu \} } [ -(1/4) \hspace{0.1 cm} \! \Psi_{\mu \nu} \Psi^{\mu \nu} ] \\[3pt] &+  \mathcal{D}^\mu A^\dagger \mathcal{D}_\mu A
   - m_{A}^2\,A^\dagger A - V(A, A^{\dagger})\\[3pt] &+ \mathcal{D}^\mu B^\dagger \mathcal{D}_\mu B - m_{B}^2\,B^\dagger B - V(B, B^{\dagger})\\[3pt] &+  \mathcal{D}^\mu C^\dagger \mathcal{D}_\mu C
   - m_{C}^2\,C^\dagger C - V(C, C^{\dagger})\\[3pt] &+ \mathcal{D}^\mu D^\dagger \mathcal{D}_\mu D
   - m_{D}^2\,D^\dagger D - V(D, D^{\dagger})\\[3pt]
&- \Big(
   g_{AB} H_2^\dagger A B^\dagger B
   + g_{BC} H_3^\dagger  B C^\dagger C
   + g_{CD} H_4^\dagger C  D^\dagger D
   + \text{h.c.}
  \Big) \\[3pt]
&+ \sum_{i=1}^{4}\!
  \Big[
     \mathcal{D}^\mu H_i^\dagger \mathcal{D}_\mu H_i
     + m_{H_i}^2\,H_i^\dagger H_i
     - \lambda_{H_i}(H_i^\dagger H_i)^2
  \Big]  
\label{eq:FullLagrangian}
\end{aligned}
\end{equation} 

Note that we use \enquote{h.c.} to denote the hermitian conjugate. The action is invariant under local transformations of the form $\Phi \to \exp({i q f(x)}) \Phi$, with $q$ denoting the charge, where we indicate the charge assignments in the table below. If we had a global symmetry, we would instead have transformations of the form $\Phi \to \exp({i q f}) \Phi$, and would not require gauge bosons, such as $\alpha_\mu$. 

\FloatBarrier

\par\medskip
\noindent
\begin{minipage}{\columnwidth}
\captionof{table}{Charge assignments for our fields under
$\mathrm{U}_1(1)\times\mathrm{U}_2(1)\times
\mathrm{U}_3(1)\times\mathrm{U}_4(1)$.}
\label{tbl1}

\centering
\begin{tabular}{@{}ccccc@{}}
\toprule
\textbf{Field}
& \(\mathbf{U_1(1)}\)
& \(\mathbf{U_2(1)}\)
& \(\mathbf{U_3(1)}\)
& \(\mathbf{U_4(1)}\) \\
\midrule
\(\mathrm{A}\)   & 0 & 1 & 0 & 0 \\
\(\mathrm{B}\)   & 0 & 0 & 1 & 0 \\
\(\mathrm{C}\)   & 0 & 0 & 0 & 1 \\
\(\mathrm{D}\)   & 0 & 0 & 0 & 1 \\
\midrule
\(\mathrm{H}_1\) & 1 & 0 & 0 & 0 \\
\(\mathrm{H}_2\) & 0 & 1 & 0 & 0 \\
\(\mathrm{H}_3\) & 0 & 0 & 1 & 0 \\
\(\mathrm{H}_4\) & 0 & 0 & 0 & 1 \\
\bottomrule
\end{tabular}
\end{minipage}
\par\medskip

\FloatBarrier

Each time a U(1) is spontaneously broken, a dark Higgs field gains a non-zero vacuum expectation value. This then sequentially turns on the terms of the form $A B^\dagger B$, $B C^\dagger C$, and $C D^\dagger D$, where both two and three-body decays are forbidden until the relevant U(1) symmetries are spontaneously broken. For a related mechanism that uses spontaneous symmetry breaking to activate decays, see~\cite{Mario_2019}. The terms $A B^\dagger B$, $B C^\dagger C$, and $C D^\dagger D$ enable decays from one field to another.

\newpage 

\section{Challenges to This Model}

Before going into the challenges in detail, we again emphasize that since we are restricting to renormalizable models, these challenges are not merely challenges faced by one specific construction, but rather generic problems one will encounter in some form, once one insists that a realization of Chen and Loeb's ideas must be a renormalizable theory. 
\subsection{Obtaining $w<0$}
Obtaining $w<0$ for the dark matter field $C$ is very challenging; this arguably constitutes the single greatest obstruction. One might try to construct a potential that is linear in the magnitude of $C$, perhaps by using a term of the form $\sqrt{C^{\dagger} C}$, as one might expect this to yield $w \approx -1/3$. The equation below yields the equation of state parameter $w(R)$ for a complex scalar field, which can be written in terms of its magnitude, $R$, and phase factor, $\phi$, giving $R \exp({i \phi})$~\citep{Boyle_2002}.
\begin{equation}
    w(R) \approx \frac{RV'(R) - 2V(R)}{RV'(R) + 2V(R)}
\end{equation}

While this may at first seem attractive, this direction faces severe challenges. Not only is $\sqrt{C^{\dagger} C}$ non-analytic at the origin and arguably unnatural, this construction cannot generate $w<0$. In order to have a negative equation of state parameter from such a potential (or, similarly, from an absolute value potential for a real scalar), one must have a coherently oscillating, homogeneous field. This is problematic when $C$ is produced via decays, as we then expect $C$ to act like a bath of particles. One could try to achieve $w<0$ from slow-roll, though this would require $C$ to behave as quintessence, rather than dark matter. One could also try to couple $C$ to dynamical dark energy, allowing for energy transfer between the two, though this would imply moving beyond the framework of Chen and Loeb, where the dynamical equation of state was due to the dark matter only, and dark energy was a cosmological constant. This severely challenges the desire to produce a consistent, natural, renormalizable realization of Chen and Loeb's framework.

\subsection{Fine-Tuning}

Another challenge faced by this framework is that there are naturally many terms which are not forbidden by the symmetries of the theory, which would generically be expected to appear with $\mathcal{O}(1)$ coupling constants. These include terms of the form $(H_1^\dagger H_1) (H_2^\dagger H_2)$,  $(A^\dagger A)^2$, and $(H_2^\dagger A) (C^\dagger C)$. We note that one can justify the absence of a coupling between the Standard Model Higgs and the dark Higgs fields here if the model is embedded in something resembling $\mathrm{E}_8 \times \mathrm{E}_8$~\citep{Witten_1985}, as it can then be natural to have a dark sector that has no non-gravitational interactions with the Standard Model. 

\subsection{Warm Dark Matter}

In order to obtain $w>0$ for some of the stages of the cascade, one needs warm dark matter, where the positive equation of state parameter comes from the decay products being produced with significant kinetic energy. Warm decay products can indeed have $w>0$, though warm dark matter is constrained by observations, e.g.~\cite{Garcia_Warm_DM_2025}, suggesting challenges.

\subsection{Contributions to Vacuum Energy}

When each dark Higgs gains a VEV, it generically contributes to the total vacuum energy density. This has the potential to result in a disastrous effect on the Hubble expansion, which is a possible additional challenge, though we note that this issue is a generic problem for many models incorporating spontaneous symmetry breaking, such as Grand Unified Theories and dark sector theories that incorporate spontaneous symmetry breaking. Even the (experimentally confirmed) Higgs boson in the Standard Model is known to produce a contribution to the vacuum energy density that is around 55 orders of magnitude greater than what we infer for dark energy~\citep{Sola_2013}, which falls under the broader cosmological constant problem. This issue has been studied for decades, with no fully satisfactory resolution. One could perhaps assume that whatever mechanism absorbs the similar vacuum energy contributions from spontaneous symmetry breaking in the Standard Model and Grand Unified Theories also absorbs the contributions from spontaneous symmetry breaking in our dark sector, though this is an unresolved issue, both for our model, and for much of particle theory in general. We thus do not consider this to be as serious a challenge as obtaining $w<0$, though we consider it worth discussing. 

\subsection{Obstructions are Generic Once Renormalizability is Required}
\label{sec:Broad_Obstruction}

One may initially object on the grounds that these difficulties seemingly only narrowly apply to the present model, though this is not the case. If one wishes to construct a renormalizable model with a few reasonable restrictions, these difficulties are in fact relatively generic. We first note two points.

\begin{enumerate}
  \item A perturbatively renormalizable theory in four spacetime dimensions requires that all operators have a mass dimension $\leq 4$.
  \item Non-polynomial operators of the form $\phi^n$, $n \notin \mathbb{N}$, do not qualify as renormalizable. One can roughly motivate this by noting that such operators can in principle be Taylor-expanded to yield an infinite series of higher-dimension operators. Operators such as $\cos(\phi)$ are non-renormalizable for similar reasons. 
\end{enumerate}

To realize the desired expansion history, we need to transition from $w_1 = 0$ to $w_2 > 0$ to $w_3 < 0$ to $w_4 > 0$. A single scalar dark matter field whose potential is of the form $V(\phi) = c_1 \phi + c_2 \phi^2 + c_3 \phi^3 + c_4 \phi^4 $ cannot achieve this. We then conclude that multiple fields are required. Even without considering the difficulties associated with achieving $w_3<0$, the transition from $w_1=0$ to $w_2>0$ cannot be achieved by evolution in a quartic potential; instead, one would expect a transition from $w_1=1/3$ to $w_2=0$~\citep{Boyle_2002}. We thus conclude that production of warm dark matter (or, more generically, some relativistic decay product) is necessary to achieve the transition from $w_1=0$ to $w_2>0$. One is then naturally led to a decay cascade once one imposes the following assumptions, many of which are just the requirement that we do not venture outside Chen and Loeb's broad framework of evolving dark matter, e.g. by introducing elements such as dynamical dark energy. 

\begin{enumerate}
  \item The theory is renormalizable and only contains polynomial interactions. 
  \item Dark energy is a cosmological constant, as found in Chen and Loeb's framework. This then forbids quintessence-like fields.
  \item The dynamical equation of state is due entirely to the dark matter, as in Chen and Loeb's framework.
  \item Gravity is described by standard General Relativity and we work in a four-dimensional spacetime.
  \item The fields in question must behave like dark matter, which implies they cannot slow-roll on their potential like a quintessence field.
  \item The oscillating equation of state is due to scalar fields, as in Chen and Loeb's work. We expect that the obstructions hold even without this strict assumption, as the introduction of fermions does not obviously alleviate the challenges, though we have not proven this rigorously. 
\end{enumerate}

\section{\label{sec:methods} Methods: Effect on Hubble Expansion}

\subsection{$H(z)$ From Our Cascade}

While our main point is that renormalizable realizations of Chen and Loeb's ideas face significant challenges, we consider it useful to provide an illustrative plot showing the effect on the Hubble expansion, if one has a decay cascade $A \to B \to C \to D$ with equation of state parameters $w_{A}, w_{B}, w_{C}, w_{D}$. The code used for conducting the following analysis is publicly available~\citep{Hoelscher2025}. We numerically integrate the following system of coupled differential equations to calculate the evolution of the Hubble parameter with cosmic time. We assume that dark matter is 85$\%$ of the total matter density at recombination, and that the dark field $A$ comprises 20$\%$ of the total dark matter at recombination. Note that $\Theta ( t - t_i)$ is a step function that becomes nonzero after $\rm{U}_{i}(1)$ is broken.  

We use equation of state parameters $w_{A} = 0$, $w_{B} = 1/12$, $w_{C} = -1/3$, and $w_{D} = 1/12$, where $B$ and $D$ are warm dark matter. While we treat these values as constants over relatively small redshift windows, this is purely a simplifying approximation, as the equation of state parameters for the warm fields would vary with redshift. One could perhaps consider $w_{C}<0$ to be a phenomenological proxy for a coupling between dark matter and dynamical dark energy~\citep{Naidoo_Lambda_e_DM}, though further consideration of that is beyond the scope of this work, as it moves beyond realizations of Chen and Loeb's ideas. Our plot for $\Delta H/ H$ should be regarded as illustrative only. 

We use $\Gamma_{A} = \Gamma_{B} = \Gamma_{C} = 10^{-15}$ $\rm{sec^{-1}}$. (See the Appendix for the equation that expresses the decay rate in terms of the masses and coupling constants.) In the equations below, $\rho_{A}$, $\rho_{B}$, $\rho_{C}$, and $\rho_{D}$ are the densities of the complex scalars in the cascade, whereas $\rho_{O}$ is the density of the other matter (baryons+other dark matter), $\rho_r$ is the radiation density, and $\rho_{\Lambda}$ is the dark energy density. In the equations below, $t_2$, $t_3$, and $t_4$ correspond to redshifts of 3, 1.5, and 0.6. We neglect any energy transfer to the massive radial modes of the dark Higgs fields via three-body decays, or annihilation after the relevant U(1) symmetries are broken; a more detailed analysis could consider this.  

\begin{subequations}
\label{eq:cascade}
\begin{align}
\dot{\rho}_{ A}+3H(1\!+\!w_{ A})\rho_{ A}
   &=-\Theta(t\!-\!t_2)\Gamma_{ A}\rho_{ A}
   \label{eq:rhoA}\\[3pt]
\dot{\rho}_{ B}+3H(1\!+\!w_{ B})\rho_{ B}
   &=\Theta(t\!-\!t_2)\Gamma_{ A}\rho_{A}
     \nonumber\\
   &\quad-\Theta(t\!-\!t_3)\Gamma_{B}\rho_{B}
   \label{eq:rhoB}\\[3pt]
\dot{\rho}_{C}+3H(1\!+\!w_{C})\rho_{C}
   &=\Theta(t\!-\!t_3)\Gamma_{B}\rho_{B}
     \nonumber\\
   &\quad-\Theta(t\!-\!t_4)\Gamma_{C}\rho_{C}
   \label{eq:rhoC}\\[3pt]
\dot{\rho}_{D}+3H(1\!+\!w_{D})\rho_{D}
   &=\Theta(t\!-\!t_4)\Gamma_{C}\rho_{C}
   \label{eq:rhoD}
\end{align}
\end{subequations}
\begin{equation}
H^{2}=\frac{8\pi G}{3}
   (\rho_{A}+\rho_{B}+\rho_{C}+\rho_{D}
   +\rho_{O}+\rho_{r}+\rho_{\Lambda})
\label{eq:friedmann}
\end{equation}

We integrate the system of equations from recombination at $z_{rec}=1090$ to $z=0$, where we specify our initial conditions by setting $h_i$ = 0.674, where $H_0$ is $100 \hspace{0.1 cm} h_i \hspace{0.1 cm} \rm{km \hspace{0.1 cm} sec^{-1} Mpc^{-1}}$ in units of $\rm{sec}^{-1}$. The radiation $\Omega_r$ and matter $\Omega_m$ components are dictated by the equations below~\citep{Dodelson, Planck_H0}, where $\Omega_\Lambda$ follows from flatness: 
\begin{equation}
    \Omega_r = 4.15 \times 10^{-5} /h_i^2
\end{equation}
\begin{equation}
    \Omega_m = 0.14241/h_i^2.
\end{equation}

\noindent The densities are then determined by the following equations, where $\rho_{m,i}$, $\rho_{r,i}$, and $\rho_{\Lambda,i}$ are the total matter, radiation, and dark energy densities at recombination ($z=1090$), respectively. 
\begin{align}
\rho_{crit,0} &= \frac{3 H_0^2}{8 \pi G} \\
\Omega_\Lambda &= 1 - \Omega_m - \Omega_r \\
\rho_{m,i} &= \Omega_m \rho_{crit,0} (1 + z_{ rec})^3 \\
\rho_{r,i} &= \Omega_r \rho_{crit,0} (1 + z_{ rec})^4 \\
\rho_{\Lambda,i} &= \Omega_\Lambda \rho_{ crit,0}
\end{align}

\noindent We assume that dark matter comprises 85$\%$ of the total matter density, and that the dark field $A$ makes up 20$\%$ of the total dark matter at recombination, so we set $\rho_{A,i} = (0.85)(0.2) \rho_{m,i}$. Note that since $\rho_{m,i}$ is the total matter density, it includes both baryonic matter and dark matter, so $\rho_{A,i}$ is a component of $\rho_{m,i}$. We assume that $\rho_{B,i} = \rho_{C,i} = \rho_{D,i} = 0$. This is because at $z=1090$ these species have not yet been produced via decays. 

\subsection{$w_0 w_a$CDM}

We use the equation below to compute $H(z)$ for a Universe with cold dark matter and dynamical dark energy~\citep{Linder_2003}. We plot $H(z)$ from our model alongside $w_0 w_a$CDM, which has the equation of state parameter $w_{DE}(a) = w_0 + w_a (1-a)$ for dark energy, to show that our model displays qualitative consistency with the DESI DR2 trends. Note that $a$ is the scale factor.

\begin{equation}
    H(z) = H_0 \sqrt{\Delta},
\end{equation}

\noindent where

\begin{equation}
\begin{split}
\Delta ={}& \Omega_{m} (1+z)^3 
         + \Omega_{r} (1+z)^4 \\
         &+ \Omega_\Lambda (1+z)^{3(1 + w_0 + w_a)} 
            \exp\!\left[-\frac{3 w_a z}{1+z}\right].
\end{split}
\end{equation}

\subsection{Effect on the Expansion History}
In Figure~\ref{fig:Hubble_Param_Evolution} we plot $\Delta H/H$, the fractional change in the Hubble parameter, as compared to standard $\Lambda$CDM. This expansion history is qualitatively similar to the DESI observations as well as the $H(z)$ produced by Chen and Loeb with step-like changes in $w$. In the figure below, $\Omega_{tot} \approx 1$ at $z=0$, which avoids tension with observational constraints on curvature~\citep{Planck_H0}. 

\FloatBarrier
\begin{minipage}{\columnwidth}
    \centering
    \includegraphics[
        width=\columnwidth,
        height=0.22\textheight,
        keepaspectratio
    ]{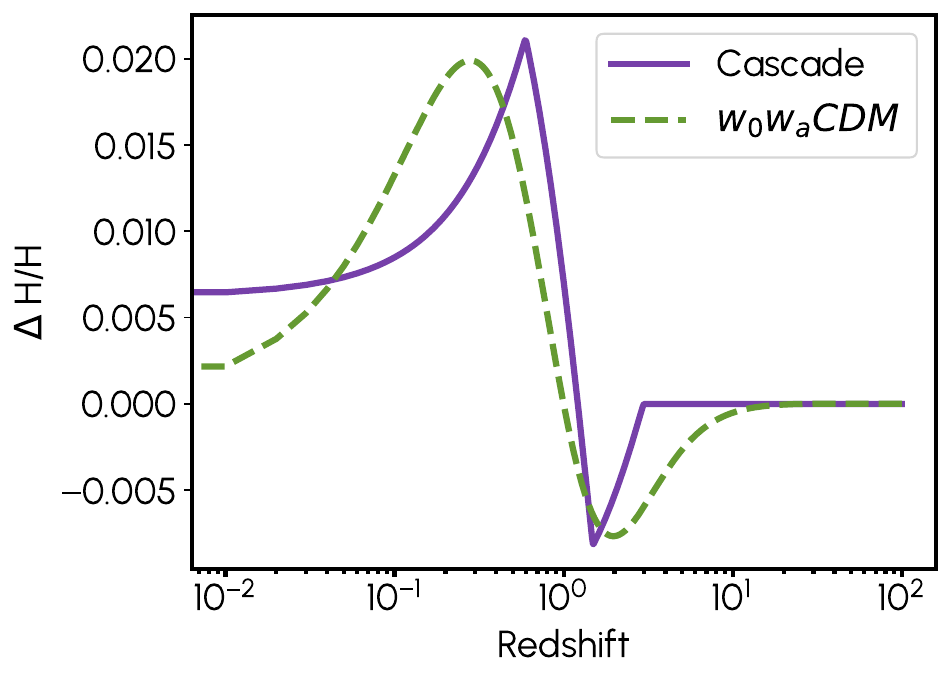}

    \captionof{figure}{The fractional change in $H(z)$ with our
    cascade, as compared to $\Lambda$CDM (solid). We also include a curve
    corresponding to dynamical dark energy with $w_0=-0.83$ and
    $w_a=-0.62$, which produces behavior similar to that favored by
    DESI (dashed).}
    \label{fig:Hubble_Param_Evolution}
\end{minipage}

\FloatBarrier

\FloatBarrier

\section{Interesting Features of our Framework}

While our primary conclusion is that renormalizable realizations of Chen and Loeb's ideas are severely constrained, we believe our framework still has some features that may be of interest for model-building. In particular, it allows for decay pathways to open at particular redshifts, while also naturally incorporating a possible self-interacting dark matter candidate. 

\subsection{Self-Interacting Dark Matter Candidate}

\FloatBarrier
After the first U(1) is spontaneously broken, yielding a dark Higgs with a mass of $\mathcal{O}$(MeV), the term $\mathcal{D}^\mu H_{1}^{\dagger} \mathcal{D}_\mu H_{1}$ produces an interaction of the form $g h \alpha_\mu \alpha^\mu$, which permits $\alpha \alpha \to \alpha \alpha$ scattering processes via a scalar mediator, as shown below. We note that this U(1) must be broken early for the fields to have appropriate masses. We use $h$ to represent the radial mode of the dark Higgs, after spontaneous symmetry breaking gives it a mass. For $\alpha \sim \rm{GeV}$ and $h \sim \rm{MeV}$, this acts as a natural SIDM candidate, one that could conceivably address some of the issues with CDM if $\alpha$ comprises the majority of the dark matter. We show the $t$ and $u$-channel Feynman diagrams in Figure~\ref{fig:Diagrams}. We note that the contribution from the $s$-channel is subdominant relative to the $t$ and $u$-channels in the relevant velocity regime ($v < \mathcal{O}(1000$ km/sec)). Refer to the Appendix for further details.

Our model does not strictly require this SIDM candidate, though it can be very naturally embedded in our framework, given that it already requires multiple U(1)'s. For example, $H_1$ could be associated with triggering some earlier stage in the decay chain, which produces the field $A$ via decays at early times. This feature is very attractive, as SIDM can help resolve important issues involving small-scale structure, which motivates natural embeddings of viable SIDM candidates into particle theory.

\begin{center}
\begin{minipage}{\columnwidth}
\centering

\tikzset{
  vec/.style = {decorate, decoration={snake, amplitude=1.4pt, segment length=6pt}, line width=0.9pt},
  sca/.style = {dashed, line width=0.9pt},
  dot/.style = {circle, fill=black, inner sep=0pt, minimum size=3pt},
  lab/.style = {font=\small}
}
\begin{tikzpicture}[line cap=round, line join=round, thick]
  \def\R{2}
  \def\dot{1.8pt}
  \usetikzlibrary{calc}

  \path
    ({\R*cos(45)},{\R*sin(45)}) coordinate (B2)
    ({\R*cos(135)},{\R*sin(135)}) coordinate (B3)
    ({\R*cos(-45)},{\R*sin(-45)}) coordinate (B1)
    ({\R*cos(-135)},{\R*sin(-135)}) coordinate (B4);

  \node[circle,fill=black,inner sep=\dot] (Vtop)
      at (0,0.35*\R) {};
  \node[circle,fill=black,inner sep=\dot] (Vbottom)
      at (0,-0.35*\R) {};

  \draw[dotted] (Vtop) -- (Vbottom);
  \draw[vec] (Vtop) -- (B2);
  \draw[vec] (Vtop) -- (B3);
  \draw[vec] (Vbottom) -- (B1);
  \draw[vec] (Vbottom) -- (B4);
\end{tikzpicture}
\hfill
\begin{tikzpicture}[line cap=round, line join=round, thick]
  \def\R{2}
  \def\dot{1.8pt}
  \usetikzlibrary{calc}

  \path
    ({\R*cos(45)},{\R*sin(45)}) coordinate (B2)
    ({\R*cos(135)},{\R*sin(135)}) coordinate (B3)
    ({\R*cos(-45)},{\R*sin(-45)}) coordinate (B1)
    ({\R*cos(-135)},{\R*sin(-135)}) coordinate (B4);

  \node[circle,fill=black,inner sep=\dot] (Vtop)
      at (0,0.35*\R) {};
  \node[circle,fill=black,inner sep=\dot] (Vbottom)
      at (0,-0.35*\R) {};

  \draw[dotted] (Vtop) -- (Vbottom);
  \draw[vec] (Vtop) -- (B1);
  \draw[vec] (Vtop) -- (B3);
  \draw[vec] (Vbottom) -- (B2);
  \draw[vec] (Vbottom) -- (B4);
\end{tikzpicture}

\captionof{figure}{
Feynman diagrams for tree-level
$\alpha\alpha\to\alpha\alpha$ scattering via a scalar mediator $h$.
The $t$-channel diagram is shown on the left and the $u$-channel
diagram on the right.
}
\label{fig:Diagrams}

\end{minipage}
\end{center}

\subsection{SIDM Cross Section}

We show the momentum-transfer cross section as a function of velocity in Figure~\ref{fig:Momentum_Transfer}, with $m_{\alpha} =$ 1 $\rm{GeV}$, $m_h =$ 0.001 $\rm{GeV}$, and $g = 0.0175$. Note that $m_{\alpha}$ is the mass of the vector, $m_h$ is the mass of the scalar mediator, and $g$ is the effective coupling constant. We also plot the differential cross section in Figure~\ref{fig:Differential_Cross_Sect_200} and Figure~\ref{fig:Differential_Cross_Sect_1000}. It is clear that the cross section decreases with increasing relative velocity. Since $(g^2 / 4 \pi)(1 / m_\alpha m_h) \approx 0.024 << 1$, perturbation theory is an appropriate approximation. Below we define the momentum-transfer cross section, $\sigma_T$~\citep{Correa_2022}. This is more useful for SIDM than the total cross section, as it down-weights forward and backward scattering by including $(1-|\rm{cos (\theta)}|)$ in the integral. We use $\theta$ for the scattering angle, $\Omega$ for the solid angle, $\mathcal{M}$ for the matrix element, and $E_{cm}$ for the energy in the center-of-mass frame.

\begin{equation}
    \sigma_{T} = 2 \int d\Omega(1-|\cos (\theta)|) \frac{|\mathcal{M}|^2}{64 \pi^2 E_{cm}^2}
\end{equation}

We include a closed-form expression for $|\mathcal{M}|^2$ in the Appendix. We employ the tree-level cross section instead of the simpler Born approximation based on a Yukawa potential because the tree-level cross section captures physics which the Born approximation does not. The Born approximation offers a simpler method for computing a cross section from a Yukawa potential, though it neglects the $u$-channel diagram, as well as the interference between $t$ and $u$-channels~\citep{Girmohanta_2022}, which our calculation incorporates. (See Figure~\ref{fig:Diagrams} for a depiction of the $t$ and $u$-channel processes.) One can view neglecting the $u$-channel as a simplifying approximation, though it dramatically changes $d\sigma/d\Omega$. A process that includes only the $t$-channel yields a differential cross section that peaks at $\theta=0$ but not $\theta=\pi$, whereas the $t+u$-channel differential cross section peaks at both $\theta=0$ and $\theta=\pi$, producing a U-shaped plot. It is preferable to use the full tree-level cross section, when possible, even though this is more complicated to compute~\citep{Girmohanta_2022}. 

\FloatBarrier

\begin{minipage}{\columnwidth}
    \centering
    \includegraphics[
        width=\columnwidth,
        height=0.22\textheight,
        keepaspectratio
    ]{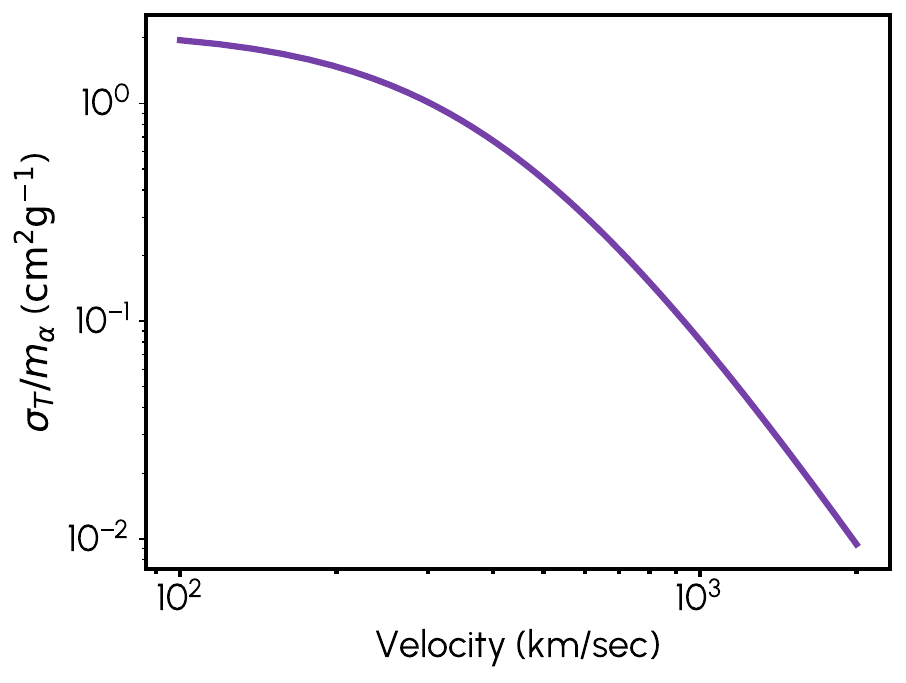}

    \captionof{figure}{We plot the tree-level momentum-transfer cross section for the scattering process $\alpha \alpha \to \alpha \alpha$ with $m_\alpha$ = 1 GeV and $m_h = 0.001$ GeV. One can see that this decreases with increasing relative velocity, potentially enabling the SIDM to evade constraints from galaxy clusters while still producing cored density profiles in galaxies.}
    \label{fig:Momentum_Transfer}
\end{minipage}

\FloatBarrier

\FloatBarrier

\begin{minipage}{\columnwidth}
    \centering
    \includegraphics[
        width=\columnwidth,
        height=0.22\textheight,
        keepaspectratio
    ]{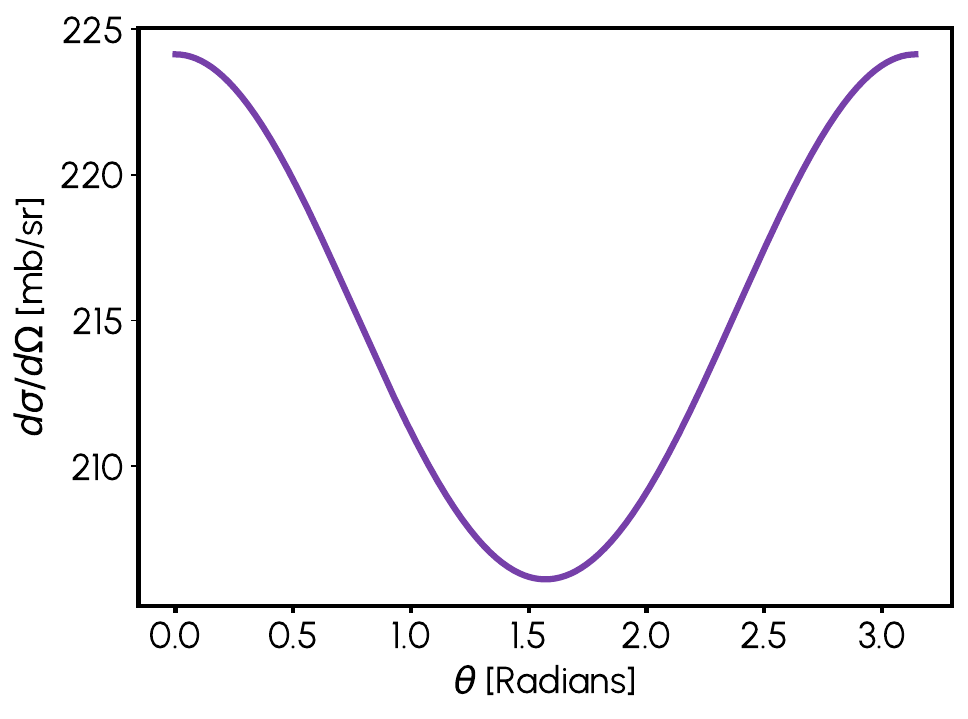}

    \captionof{figure}{We plot the tree-level differential cross section for the scattering process $\alpha \alpha \to \alpha \alpha$ with $m_\alpha$ = 1 GeV, $m_h = 0.001$ GeV, and a relative velocity of 200 km/sec. This velocity is relevant for the Milky Way scale. }
    \label{fig:Differential_Cross_Sect_200}
\end{minipage}

\FloatBarrier

\FloatBarrier

\par\medskip
\begin{minipage}{\columnwidth}
    \centering
    \includegraphics[
        width=\columnwidth,
        height=0.22\textheight,
        keepaspectratio
    ]{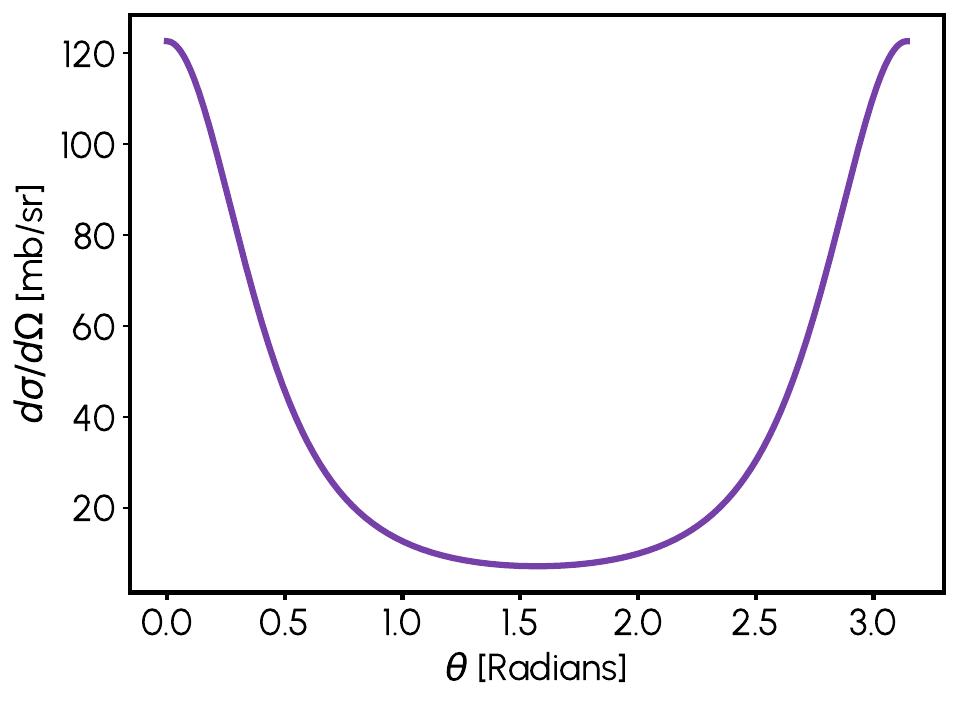}

    \captionof{figure}{We plot the tree-level differential cross section for the scattering process $\alpha \alpha \to \alpha \alpha$ with $m_\alpha$ = 1 GeV, $m_h = 0.001$ GeV, and a relative velocity of 1000 km/sec. This velocity is relevant for the galaxy cluster scale. One can see that this cross section is suppressed as compared to the result for 200 km/sec.}
    \label{fig:Differential_Cross_Sect_1000}
\end{minipage}
\par\medskip

\FloatBarrier

\section{\label{sec:conclusion}Conclusion}

While we find it to be very difficult to produce a natural, fully renormalizable model that can realize the behavior desired by Chen and Loeb, we are able to put interesting constraints on the space of such realizations, while producing a framework that has some features that may be of interest elsewhere for model-building. In particular, our framework  naturally accommodates a possible self-interacting dark matter candidate, while also allowing for decay pathways to open at particular redshifts. This study is valuable for examining the space of natural, renormalizable field theories that have the potential to realize the behavior desired by Chen and Loeb, which is relevant for finding a pathway towards explaining the DESI trends.

\section{Acknowledgements}

\noindent KHB acknowledges the support of NSF 2125764. 

\section{\label{sec:Appendix}Appendix}

\subsection{S-Matrix Amplitude for Tree-Level Scattering Cross Section}

Here we include a closed-form expression for the S-matrix amplitude used in plotting the SIDM cross section, where the relevant interaction is $g h \alpha_{\mu} \alpha^{\mu}$. Our calculation for the amplitude is publicly available as a PDF that is included with our code~\citep{Hoelscher2025}. This was produced from the $t$ and $u$-channel Feynman diagrams via a manual computation. For compactness, we define $C_1$, $C_2$, and $C_3$ as the following quantities. We use $p$ for the momentum, $E_{cm}$ for the energy in the center-of-mass frame, $\theta$ for the scattering angle, $m_h$ for the mass of the scalar mediator, and $m_{\alpha}$ for the mass of the vector field. 

\begin{align}
C_1 &= \tfrac14 E_{cm}^2 - p^2 \cos(\theta) \\
C_2 &= \tfrac14 E_{cm}^2 + p^2 \cos(\theta) \\
C_3 &= \tfrac14 E_{cm}^2 + p^2 
\end{align}

\FloatBarrier

\begin{equation}
\begin{aligned}
|\mathrm{\mathcal{M}}|^2 &= 
\frac{8 g^4}{9 m_\alpha^8}
\Bigg[
  \frac{
    4 m_\alpha^4 (C_1^2 + C_2^2 + C_3^2)
    + 2 C_1^2 C_2^2
    - 8 C_1 C_2 C_3 m_\alpha^2
  }{
    (m_h^2 + 2p^2)^2 - 4p^4\cos^2(\theta)
  }  \\[3pt]
&\qquad\qquad
 + \frac{(C_1^2 + 2 m_\alpha^4)^2}{
    (m_h^2 - 2p^2\cos(\theta) + 2p^2)^2
  } \\[3pt]
 &\qquad\qquad + \frac{(C_2^2 + 2 m_\alpha^4)^2}{
    (m_h^2 + 2p^2\cos(\theta) + 2p^2)^2
  }
\Bigg]
\label{eq:M2}
\end{aligned}
\end{equation}

\FloatBarrier

\subsection{Justification for Neglecting the $s$-Channel for SIDM}

\noindent We define the Mandelstam variables as shown below:
\begin{equation}
    s = (2 E)^2 = E^2_{cm},
\end{equation}
\begin{equation}
    t = -2p^2(1-\cos(\theta)),
\end{equation}
\begin{equation}
    u = -2p^2(1+\cos(\theta)).
\end{equation}

\noindent For $v <<< 1$, $E_{cm}^2 \approx 4 m_{\alpha}^2$ and $p \approx m_{\alpha} v$. Note that $m_{\alpha} >>> m_{h}$. This suggests:
\begin{equation}
    \biggl|\frac{1}{s - m_{h}^2}\biggr| <<< \biggl|\frac{1}{t - m_{h}^2}\biggr|,
\end{equation}
\begin{equation}
    \biggl|\frac{1}{s - m_{h}^2}\biggr| <<< \biggl|\frac{1}{u - m_{h}^2}\biggr|.
\end{equation}

\noindent While the $s$-channel exists, it is suppressed in the velocity regime we are interested in for dark matter halos (relative velocities $< \mathcal{O}$ (1000) km/sec). This justifies considering only the $t$ and $u$-channel diagrams for the dominant contributions, when computing the tree-level cross section.

\subsection{Tree-Level Decay Rate}

After spontaneous symmetry breaking, we get interaction terms of the form $\frac{g_{AB} \nu}{\sqrt{2}} A B^\dagger B$, where $\nu$ is the VEV (vacuum expectation value) of a dark Higgs. The decay rate for $A \to B^\dagger B$ can be computed from the particle masses, $m_{A}$ and $m_{B}$, and the coupling constant, $g_{AB}$, as shown below, with $c=1$, $\nu$ in units of eV, masses in units of eV, $\hbar \approx 6.58 \times 10^{-16}$ eV $\cdot$ sec, and $\Gamma$ in units of $\rm{sec}^{-1}$. The viable mass range for $A$ (and by extension, $B$, $C$, and $D$) depends in part on the production mechanism for the dark matter. 

\begin{equation}
    \Gamma_{A \to  B^{\dagger} B} = \frac{g_{AB}^2 \nu^2}{16 \pi \hbar m_{A}} \sqrt{1-\frac{4 m_{B}^2}{m_{A}^2}}
\end{equation}

In our analysis, we neglect any energy transfer from $A$ to $h$ via $\frac{g_{A B} h}{\sqrt{2}} A B^\dagger B$, where $h$ is the massive radial mode of the dark Higgs after spontaneous symmetry breaking. A more detailed analysis could consider this for the decay of $A$, as well as similarly for decays of $B$ and $C$. 

\bigskip 
\section{Declaration of Generative AI and AI-assisted Technologies in the Manuscript Preparation Process} During the preparation of this manuscript, the authors used ChatGPT as a sounding board while refining arguments (e.g. Section~\ref{sec:Broad_Obstruction}) and for occasional assistance with LaTeX. The models, key ideas, calculations, code, and writing in this manuscript were produced by the authors, rather than AI tools. 

\printcredits

\bibliographystyle{cas-model2-names}

\bibliography{cas-refs}



\end{document}